\documentclass[5p,twocolumn]{elsarticle}
\usepackage{lcy}
\usepackage{amssymb}
\usepackage{graphicx}
\begin{document}
\title{Study of the contribution of long-wave bending vibrations to the destruction of 
ultrathin films by the method of molecular dynamics
}
\author{E.I. Salamatov, E.B. Dolgusheva}
\date{}
\ead{elena@udman.ru}
\address{Udmurt Federal Research Center of Ural Branch of the Russian Academy of Sciences,
34, T.Baramzina Str., 426067, Izhevsk, Russia}
\date{\today}

\begin{abstract}

The molecular dynamics method is used to study the process of development of dynamic 
instability of a thin film, leading to its destruction. The calculations are performed for a thin 
(5 atomic layers) $fcc$ aluminum film using the interatomic interaction potential tested by comparing
the numerical results with the analytical ones obtained in the framework of elasticity theory. 
For this purpose, an original approach is developed, which allows one to calculate the dispersion 
law of long-wave phonons in ultrathin films using the molecular dynamics method. The temperatures 
($< 600 K$) at which the system remains stable over a time interval of $0.6 ns$ are found. This makes it
possible to analyze the low-frequency part of the spectrum down to the minimum frequency $\nu_{min}=0.0166
THz$ (at $T = 50 K$), and to determine the vibration frequency of the longest, for this problem geometry,
bending wave $\nu_{0}=0.033 THz$ which decreases with increasing temperature, hence, its period grows. 
Once the vibration period of this mode becomes comparable with the time of simulation, there occurs,
during calculation, a continuous increase in the amplitude of this mode which will be referred to as
``retarded mode''. It is shown that the film destruction begins with the attainment of
a certain critical value of the bending wave amplitude.
 
\textit{Keywords: Molecular dynamics method; metastable state; Fourier transform;
dispersion law of thin-film vibrations; ``retarded mode''. }

\end{abstract}

\maketitle

\section {Introduction}

Recently, much attention has been paid to the study of metastable nanofilms with unique properties that 
are finding ever new industrial applications. One of the main tasks facing the researchers 
is to increase their stability, since  numerous experiments on the synthesis of ultrathin (of several 
atomic layers) films show that such films in the free state are unstable: they curl up, bend or even 
collapse \cite{1,2,3,4}. To solve this problem, it is necessary to understand which 
changes in the system are indicative of a loss of stability, and to discover the factors responsible 
for this process.

When describing structural transitions in bulk crystals, as a rule, the concept of dynamic stability 
is used, according to which the dynamic matrix of a stable phase should be positively definite \cite{5}.
When moving to the phase boundary, at some point of the Brillouin zone, there arises a phonon mode 
whose frequency tends to zero in approaching the interface, and the dynamic matrix ceases to be positive 
definite. Atomic displacements in this phonon mode describe the process of phase transition at the microscopic 
level in real space \cite{6, 7}. This approach is confirmed by experimental data obtained by the methods of neutron 
scattering \cite{8} and Raman spectroscopy \cite{9}. The lifetime of thin crystalline films in the free state is too short 
for such experiments. One could suppose that for small times it is promising to use the method of molecular 
dynamics (MMD) which makes it possible to consider a detailed microscopic picture of the movement of large 
atomic systems on small time intervals. However, the MMD calculations must first be tested by comparing them 
with the known physical characteristics of the modeled objects, in order to make sure of the description 
validity. When simulating the thermodynamic properties of bulk systems, the calculated and experimentally 
determined density of phonon states are usually compared. We think that in the absence of experimental data,
testing can be carried out by a comparison with the results obtained in other widely-accepted approaches.
In this work, a method is suggested that allows testing the results of MD calculations of the thin film dynamics
by comparing them with the analytical results obtained in the framework of the elasticity theory. Since in the 
elasticity theory the dynamics of a crystalline system is described only in the long-wavelength limit, this
approach does not require calculating the density of phonon states, but it calls for a more thorough analysis
of the low-frequency portion of the vibrational spectrum of individual atoms of the system. The paper presents 
an original technique for calculating the long-wavelength region of the phonon dispersion law in thin films by 
the MD method, which is based on a preliminary analysis of the vibration symmetry of the film atoms.

The performed symmetry analysis makes it possible to unambiguously establish to which branch of the phonon spectrum belong 
the considered vibrations of frequency $\nu$, and to  find the wavelength $\lambda$ of the corresponding phonon, which
allows the dispersion law $\nu (k)$ to be determined. The dispersion law obtained in this way for long-wavelength
phonons of a thin aluminum film containing only five atomic monolayers coincides with the analytical result
from the theory of elasticity. This suggests that the MD method, statement of the problem, and the chosen 
interatomic potential can adequately describe the dynamics of ultrathin aluminum films. Unfortunately, the 
MD method is limited by the computation time, the size of the calculated cell, and the imposed boundary 
conditions. Under periodic boundary conditions in two directions necessary for calculating the dynamic properties
of the film, it is impossible to obtain real destruction of the film: rupture or curling up. Therefore,
by destruction we will mean such changes in the system that lead to a program halt, assuming that the same changes 
occur in the real case at the initial stage of the destruction process.
In this paper, the reasons for the development of dynamic instability of a thin film leading to its destruction 
are determined, which made it possible to suggest a reasonable scenario of this process at the atomic level.

\section{Calculational methods}

To describe the interatomic interaction in aluminum, a many-particle potential \cite{10} constructed in the ``embedded atom
model''\cite{11} was chosen. The authors of [10] showed that this potential makes it possible to obtain, to a high degree of
accuracy, the parameters of $fcc$ aluminum: cohesive energy, elastic constants, melting temperature, and other physical 
characteristics of bulk aluminum. Earlier, in Refs. \cite{12,13}, this potential was used to obtain the vibrational densities 
of states at different temperatures, elastic moduli, temperature dependence of heat capacity, thermal expansion, etc. for
$fcc$-Al, both in the bulk state and nanofilms. A comparison of the calculated characteristics with the experiment on inelastic 
neutron scattering for Al shows that the potential chosen allows one to describe the experimentally observed features of 
the aluminum phonon spectrum, including its ``softening'' with increasing temperature.

\begin{figure}[tbh]
\begin{center}
\resizebox{0.95\columnwidth}{!}{\includegraphics*[angle=-0]{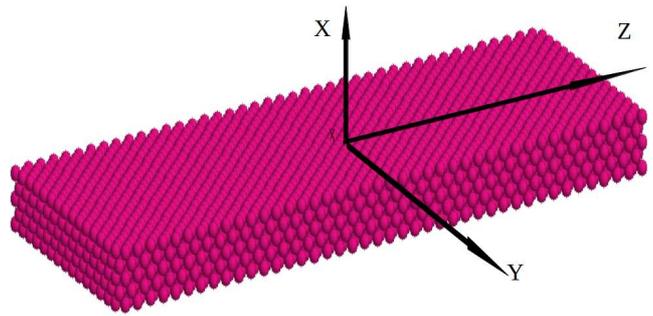}}

\caption
{The modeled base crystallite imitating a film. The film thickness $N_{x} = 5$ monoatomic Al layers, $N_{y}=30, N_{z}=84$. 
There are $6300$ atoms in all.}

\label{Fig1}
\end{center}
\end{figure}

Simulation was carried out using the XMD package \cite{14}. In all cases, the calculation began with the initiation of 
a crystallite with 
an ideal $fcc$ structure, relaxation for $0.1 ns$ under conditions of NPT-ensemble, minimization of the energy at a given
temperature and 
zero pressure, and the cell parameters determination. The calculations were performed on a crystallite with dimensions:
$N_{x}=5, N_{y}=30, 
N_{z}=84$ monoatomic layers. For a $fcc$ lattice parameter of Al $a=3.976$ \AA{} , the crystallite sizes  $L_{x},  L_{y}$ 
and $L_{z}$ are, respectively, $0.994 \times 5.964 \times 16.6992 nm$. This crystallite, shown in Fig.\ref{Fig1}, 
can be represented as 150 atomic chains of 42 atoms each, along the $z$ axis,
a total of 6300 atoms. An odd number of layers in the $x$ direction makes it easy to go to symmetrized coordinates. 
The time step was $\Delta t = 0.3fs$. Periodic boundary conditions were set along the $y$ and $z$ axes, 
and free ones along the $x$ axis. The speed scaling regime was 
used to set a constant temperature, and the Berendsen barostat to maintain a constant pressure.

\section{Determination of the thin film lifetime} 

\begin{figure}[tbh]
\begin{center}
\resizebox{0.95\columnwidth}{!}{\includegraphics*[angle=-90]{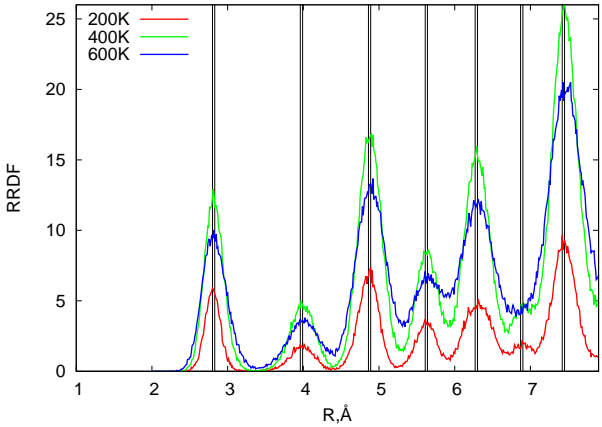}}
\caption{Functions of the radial distribution of Al film atoms at various temperatures. Straight double lines show the 
peak positions for an ideal $fcc$ lattice.}

\label{Fig2}
\end{center}
\end{figure}

Before conducting a Fourier analysis of the frequency spectrum of the system under study, it is necessary to determine the time 
interval 
($Dt$) during which the system remains stable at all considered temperatures. $Dt$ should be large enough, since the minimum 
Fourier transform 
frequency is related to the observation time by the expression $\nu_{min}=1/Dt$. An optimal time interval $Dt = 0.6ns$ was 
selected, which allows one 
to analyze the vibrational spectrum with a minimum frequency $\nu_{min}=0.0166THz$.
Figure \ref{Fig2} shows the functions of radial distribution of the film atoms at various temperatures. The double vertical 
lines indicate the atom 
positions corresponding to the ideal $fcc$ structure. It can be seen from the figure that in the temperature range 
$200 \textendash 600 K$, the lattice retains
its initial structure, and the base crystallite sizes remain practically unchanged.

\begin{figure}[tbh]
\begin{center}
\resizebox{0.95\columnwidth}{!}{\includegraphics*[angle=-90]{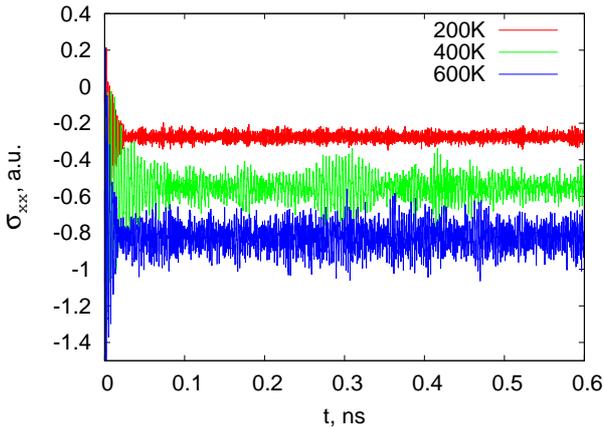}}
\caption{The time dependence of the internal stress tensor element $\sigma_{xx}$ at various temperatures.}

\label{Fig3}
\end{center}
\end{figure}

In addition to the radial distribution function, changes in the tensor of internal stress were monitored. When calculating
the elements of this tensor,
the virial contribution was not considered, and only terms due to the interatomic interaction forces were taken into account. 
The diagonal elements of the tensor are the ratio between the sum of the force projections onto the normal and the area of 
the considered
crystallite face. In this temperature range, the diagonal elements of the internal stress tensor, $\sigma_{yy}$ and $\sigma
_{zz}$  are equal to zero and practically
do not change with temperature, while the element $\sigma_{xx}$, responsible for the maintenance of the film shape,
is negative, and its modulus increases
with temperature, as shown in Fig.\ref{Fig3}.

Thus, the temperature ($0$ - $600 K$) and time ($0.6 ns$) intervals of the system stability were determined, which made it possible 
to perform a Fourier analysis of the vibrations for frequencies $\nu \geq \nu_{min}=0.0166 THz$.

\section{The method of calculating the low-frequency dispersion spectrum}

\begin{figure}[tbh]
\begin{center}
\resizebox{0.95\columnwidth}{!}{\includegraphics*[angle=-0]{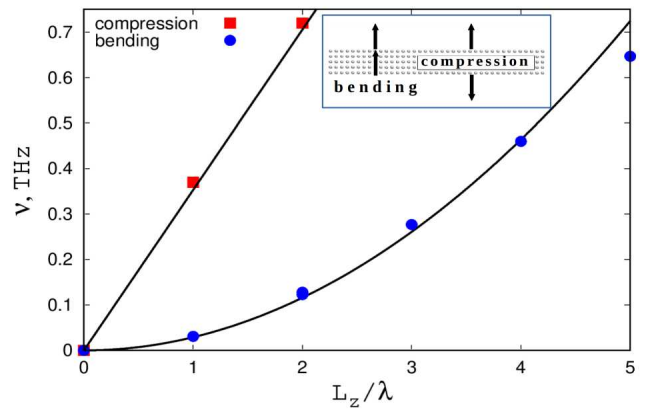}}
\caption{Dispersion of the long-wavelength phonons propagating along the $z$ direction.
Solid lines denote the calculation by formulae (1, 2) with parameters for a bulk crystal from \cite{16}.
The symbols are the results of our calculation. A schematic representation of bending and compression vibrations is shown in the inset.}

\label{Fig4}
\end{center}
\end{figure}

As already mentioned in Introduction, there are no experimental data on the dynamic properties of free ultrathin films.
The results of numerical calculations of the dispersion law can only be compared with the analytical ones obtained in the elasticity 
theory which describes well the behavior of a discrete crystalline lattice in the long-wavelength limit.      
From the theory of elasticity it follows that the presence of a plane of symmetry parallel to the surface and passing through 
the middle of the film thickness, allows one to represent the displacements of all atoms as the sum of the symmetric ($c-compression$) 
and antisymmetric ($b-bending$) contributions (see the inset in Fig \ref{Fig4}). The dispersion laws for these types of vibrations are
fundamentally different and can be calculated by the formulae from Ref. \cite{15} with the Al elastic parameters\cite{16}:
\begin{equation}
{\qquad \nu_{b}(k) = (((1+\mu)/6)^{1/2} \cdot c_{t} \cdot h_{0} \cdot k^{2} )/\pi},
\end{equation}

\begin{equation}
{\qquad \qquad \nu_{c}(k) = c_{t} \cdot k /2\cdot\pi },
\end{equation}

where $k$ is a wave vector, $c_{t}$ is the velocity of the transverse sound wave in the bulk material, $\mu$ is the Poisson's 
ratio, $h_{0}$ is the film thickness. 
The results of these calculations are shown by solid lines in Fig.\ref{Fig4}

When constructing the dispersion curves for these types of vibrations, we can restrict ourselves to considering the vibration
projections
onto the $x$ axis of two chains of surface atoms with the same value of the $y$ coordinate. The odd number of layers 
in the $x$ direction makes 
it easy to go from the real coordinates of the displacements to the symmetrized ones:

\begin{equation}
\begin{array}{c}
{\qquad \qquad u_{n,b} = (u_{n,top} + u_{n,bot})/2};\\
{\qquad \qquad u_{n,c} = (u_{n,top} - u_{n,bot})/2}, 
\end{array}
\end{equation}
where the index n = 1, ..., 42 denotes the atom number in the chain, and $top (bot)$ indicates the chain on
the upper or lower surface layer of the film. The way of finding the values designated by symbols on these curves will 
be discussed below.

\begin{figure}[tbh]
\begin{center}
\resizebox{0.95\columnwidth}{!}{\includegraphics*[angle=-90]{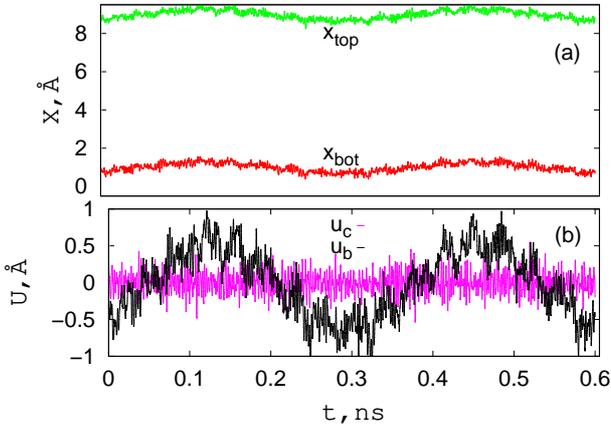}}
\caption{The time dependence of the $x$ coordinate of a pair of atoms from the upper and lower surface atomic chains
having 
the same coordinate $y = 28.7$\AA{} and the consecutive number $n = 20$ along the $z$ axis at a system temperature
$T = 200 K$  (a);
the time changes in the symmetrized displacements $u_{c}$  and $u_{b}$  (b).}
\label{Fig5}
\end{center}
\end{figure}

Figure \ref{Fig5}(a) presents the time dependence of the $x$ coordinate of a pair of atoms from the upper and lower
surface atomic chains having 
the same coordinate $y = 28.7$ \AA{} and the number $n = 20$ along the $z$ axis at a system temperature 
$T = 200 K$. Figure \ref{Fig5}(b) shows 
the corresponding changes in the symmetrized displacements $u_{c}$  and $u_{b}$. It follows from the figure that the
atomic displacements perpendicular 
to the film surface are mainly determined by the bending vibrations $u_{b}$.

\begin{figure}[tbh]
\begin{center}
\resizebox{0.92\columnwidth}{!}{\includegraphics*[angle=-0]{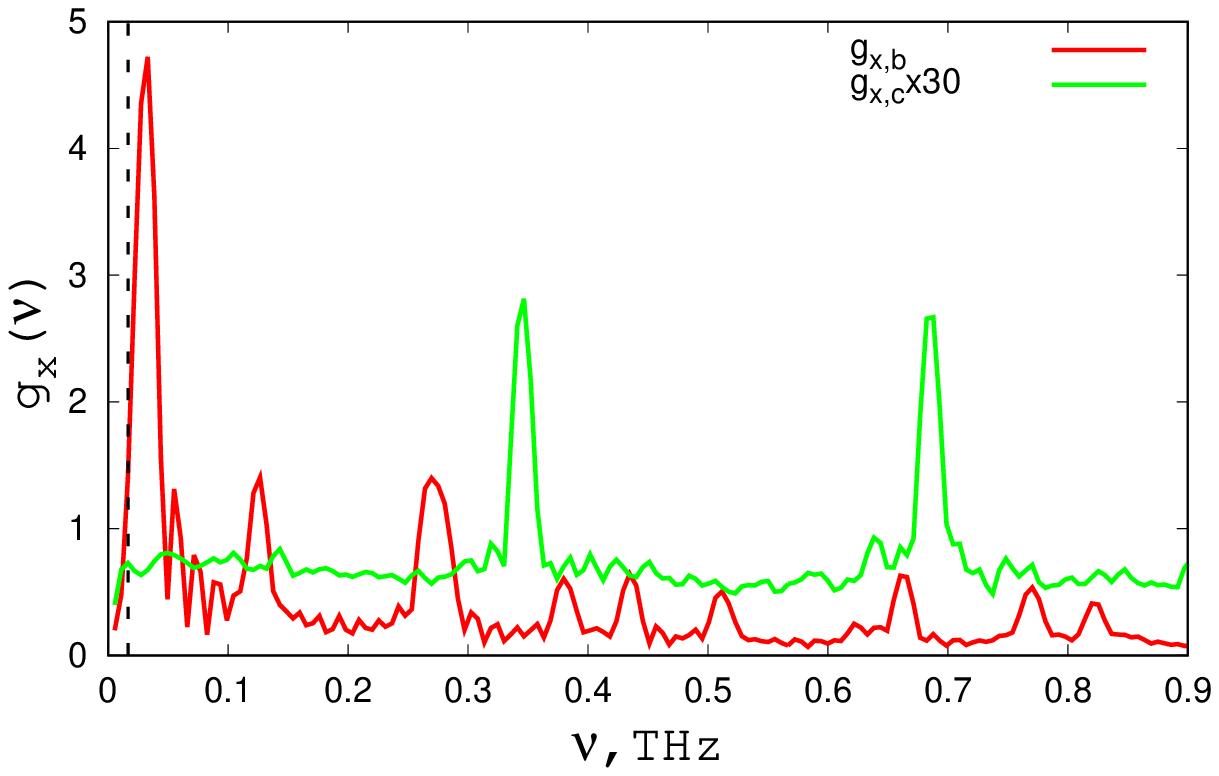}}
\caption{Low-frequency spectra of bending ($g_{x,b}$) and compression ($g_{x,c}$) vibrations of surface atoms.
The spectrum of compression vibrations  is increased 30 times. The dashed vertical line shows the minimum 
frequency $\nu_{min} = 0.0166 THz$.}
\label{Fig6}
\end{center}
\end{figure}

The Fourier transform ($F (\nu))$ of the $b, c$ - type displacement trajectories was performed for each value of $n$ 
in the chains selected. 
In this case, the frequency spectra (i.e., the Fourier transform modulus of the bending ($g_{x,b}$) and compression 
($g_{x,c}$) atomic vibrations) 
are the sum of the spectra of individual pairs of atoms from the upper and lower chains with the same y coordinate. 
The low-frequency 
portion of these spectra is depicted in Fig.\ref{Fig6}, where resonance peaks are clearly visible. The spectrum of 
bending vibrations is shown
by the red line, and that of compression vibrations (increased by $30$ times) by the green one. The dashed vertical 
line shows the minimum 
frequency $\nu_{min}=0.0166 THz$.

\begin{figure}[tbh]
\begin{center}
\resizebox{0.95\columnwidth}{!}{\includegraphics*[angle=-90]{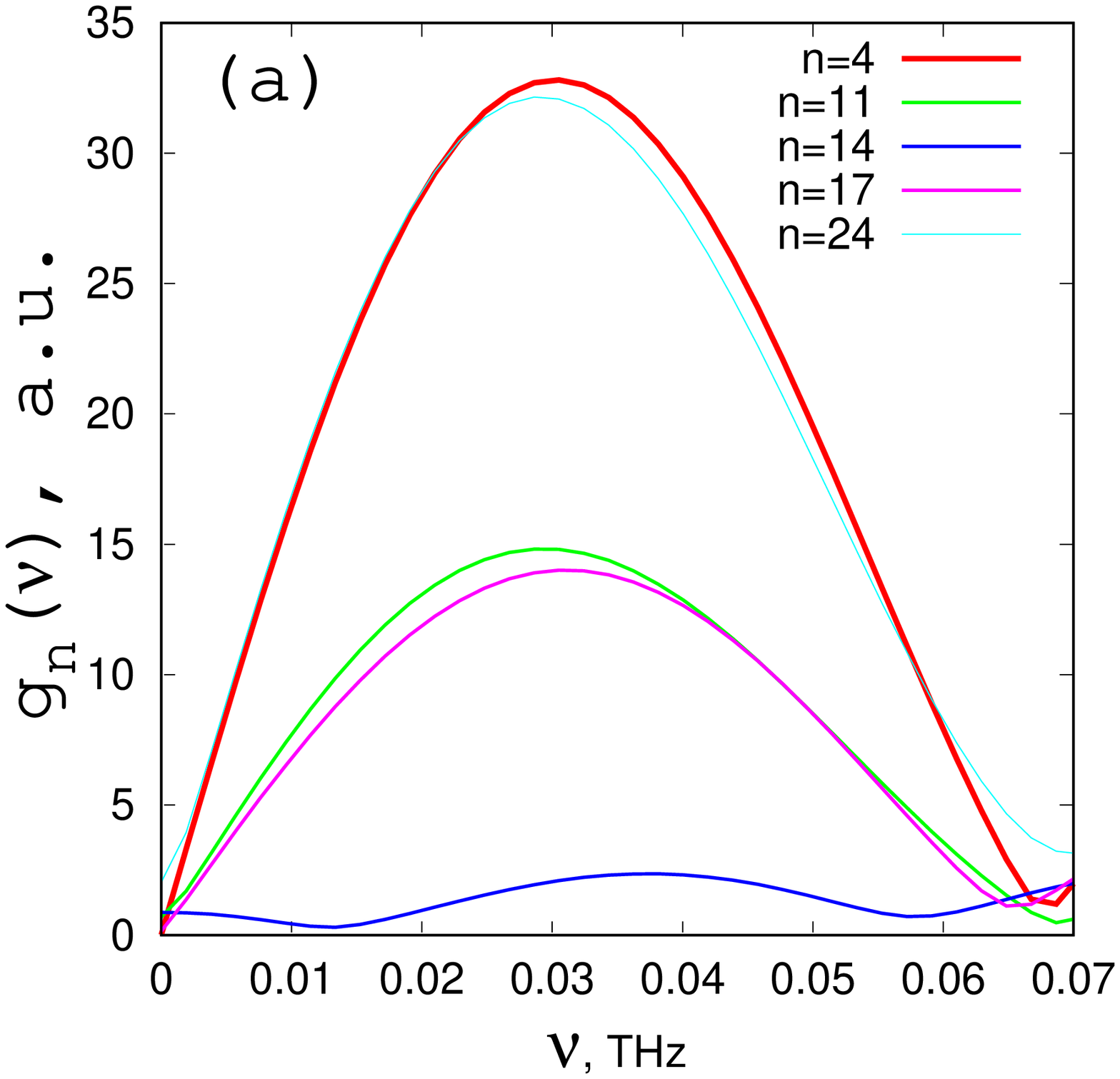}}
\resizebox{0.95\columnwidth}{!}{\includegraphics*[angle=-90]{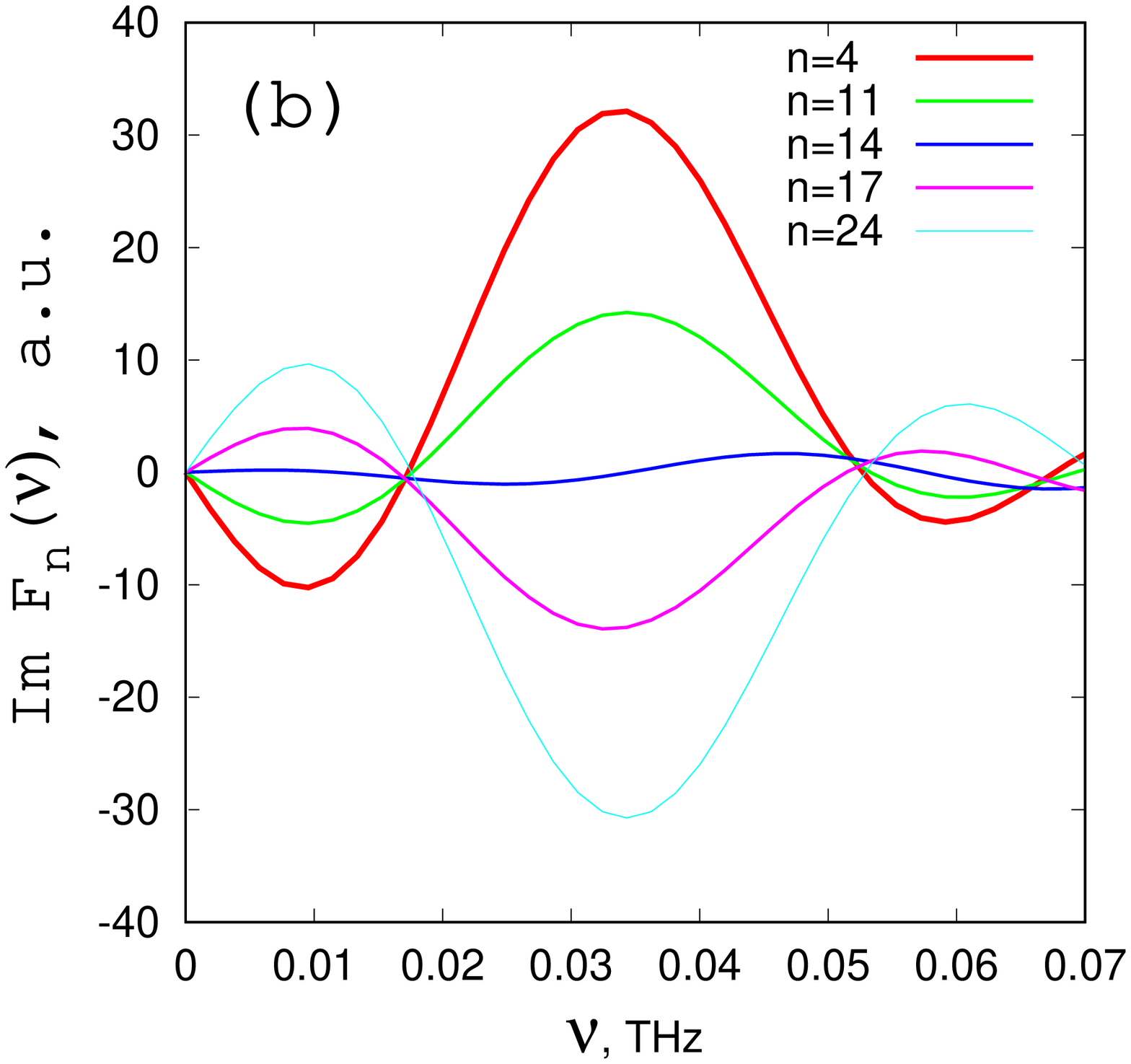}}
\caption{The contributions to the resonance frequency $\nu_{0}=0.033THz$ of the spectrum of the $n$th atomic pair  (a).
The imaginary part of the Fourier transform of bending vibrations at the same frequency  (b). 
$n$ is the atom number in the chain.}
\label{Fig7}
\end{center}
\end{figure}

Each resonance peak of the spectrum is related to a vibrational mode. To factorize it, one should consider 
the contributions from each pair of atoms to the given resonance peak. This is illustrated in Fig.\ref{Fig7}(a)
that shows the contributions
of some pairs ($n$ is the pair number along the $z$ direction) to the spectrum $g_{n}$ (modulus of $F_{n}(\nu)$) 
near the resonance with
a frequency $\nu_{o}=0.033THz$  related to bending vibrations. The frequency spectrum is a positively definite
quantity, 
and the imaginary part of the Fourier transform ($Im(F_{n}(\nu)$) saves information about the vibration phase
(see Fig.\ref{Fig7}(b)).

\begin{figure}[tbh]
\begin{center}
\resizebox{0.9\columnwidth}{!}{\includegraphics*[angle=-90]{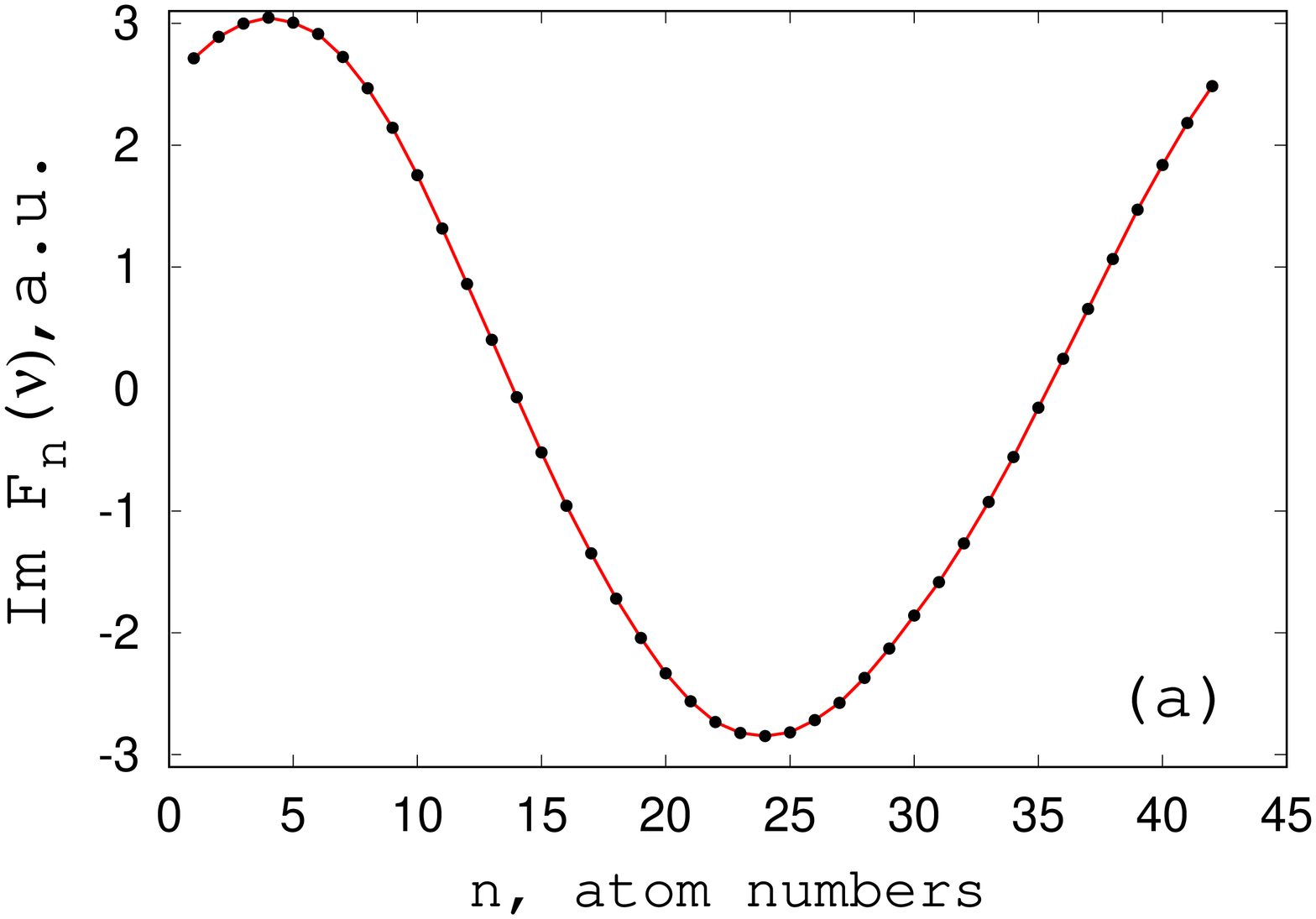}}
\resizebox{0.9\columnwidth}{!}{\includegraphics*[angle=-90]{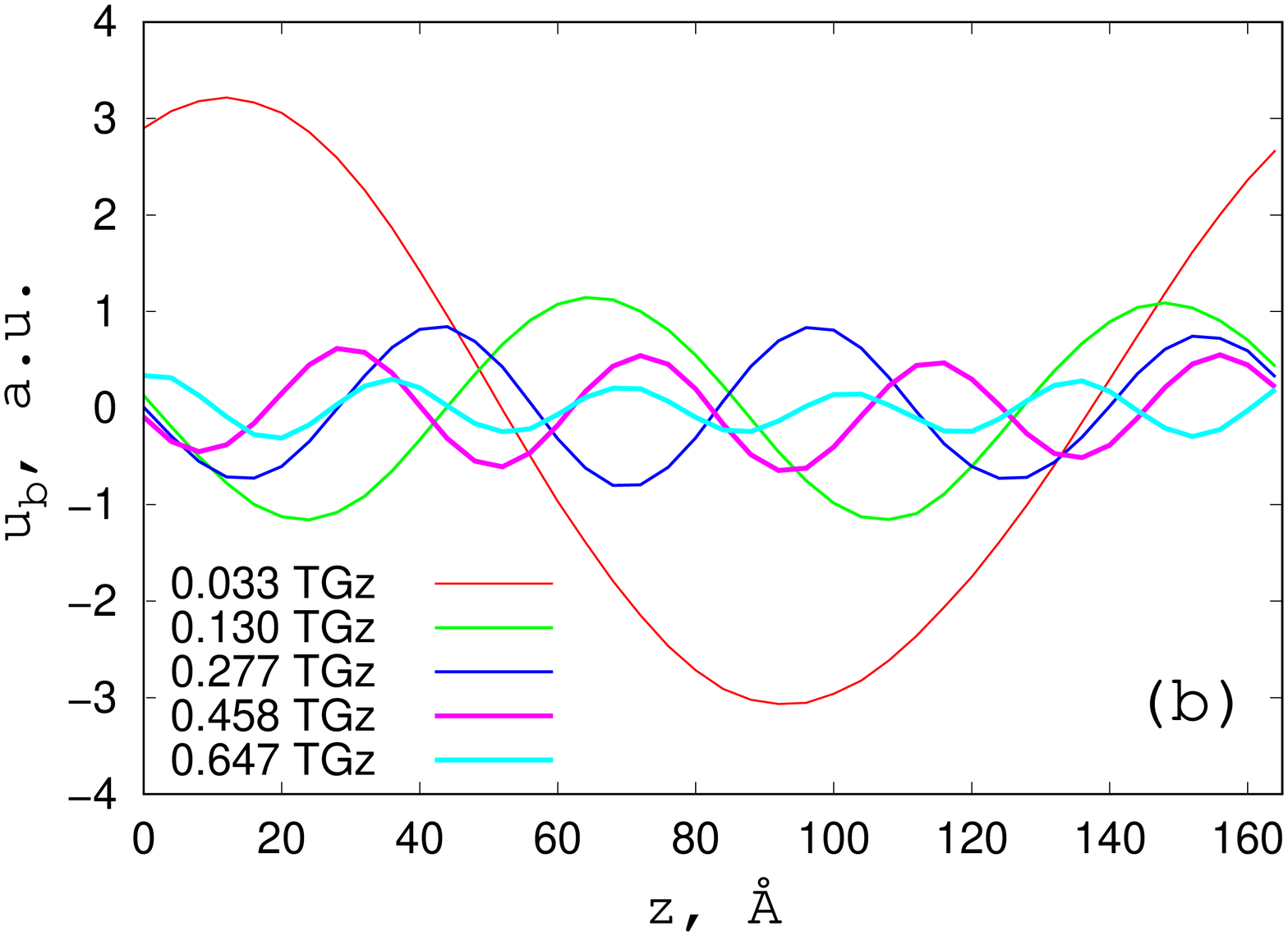}}
\caption{The imaginary values of the Fourier transform $Im F_{n}(\nu)$ at the resonance frequency $\nu_{0}=0.033THz$ 
for the entire atomic chain, 
indicated by the points (a); the bending vibrations corresponding to different resonance frequencies (b).}
\label{Fig8}
\end{center}
\end{figure}

Fig.\ref{Fig8} demonstrates the imaginary values of the Fourier transform  $Im F_{n}(\nu)$ at the resonance
frequency $\nu_{0}=0.033THz$ for the entire atomic chain,
indicated by the points.  Their envelope is a bending wave in real space, propagating along the $z$ direction;
the wavelength $\lambda$ equals the size
of the calculated cell in the $z$ direction, being maximum for the given problem geometry.

Applying the same procedure for higher-frequency resonance peaks in the spectrum of bending vibrations
(see Fig.\ref{Fig6}), the relationship between 
the wavelength and frequency was found, i.e., the dispersion law for bending vibrations was calculated.
Figure \ref{Fig8}(b) shows the bending vibrations corresponding to different resonance frequencies; 
for each of these a wavelength can be determined. 
Similarly, relationship between the resonance frequencies and wavelengths can be found for compression vibrations. 
The values obtained in this way 
correspond to the symbols in Fig.\ref{Fig4}.  Good agreement between the calculated results and the analytical 
ones obtained in the elasticity theory 
points to the validity of the approach in studying the low-frequency portion of the film vibrational spectrum,
which shows that this approach can be used to describe the beginning of the film destruction.

\section{FILM DESTRUCTION}

\begin{figure}[tbh]
\begin{center}
\resizebox{0.9\columnwidth}{!}{\includegraphics*[angle=-0]{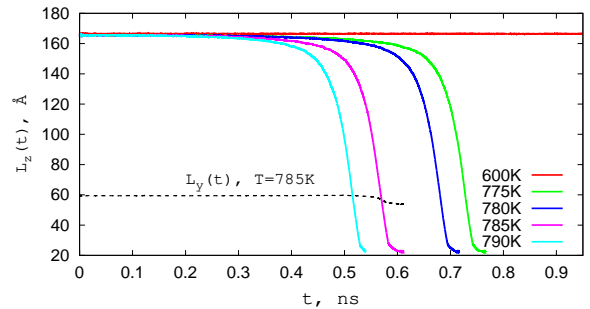}}
\caption{ The time dependence of the crystallite size $L_{z}(t)$ along the $z$ axis, obtained at various 
temperatures. $L_{y}(t)$ is obtained for $T = 785 K$.}
\label{Fig9}
\end{center}
\end{figure}

This section presents the results of investigation of the film dynamic instability development, leading to its 
destruction at high 
temperatures. As was already mentioned in  Introduction, the peculiarities of the MD method associated with the 
computation time, arbitrary choice
of the size of the calculated cell, and the imposed boundary conditions do not make it possible to obtain a real 
film destruction: its rupture 
or curling up. Therefore, by ``destruction'' we mean such changes in the system that lead to a program halt, 
assuming that the same 
changes occur, at the initial stage of destruction, in real situation.
As the temperature approaches $800 K$, the time of life (equilibrium state) of the film becomes less than $0.6~ns$.
This manifests
itself most clearly by a rapid exponential decrease in the size of the base crystallite along the $z$ direction,
$L_{z}(t)$, at times greater than a 
certain critical value $t_{0}$. Note that the time $t_{0}$ is not only a function of temperature, it also depends on 
how the system is brought to a 
given temperature. All calculated time dependences of the crystallite size $L_{z}(t)$ shown in Fig.\ref{Fig9}
started after
the system was kept for $0.1 ns$ at $300 K$, which allows one to assume that the conditions of reaching high 
temperatures ($> 770 K$) are  almost the same for all $L_{z}(t)$ curves. Nevertheless, it is seen 
from Fig.\ref{Fig9} that the expected decrease in the parameter $t_{0}$ with increasing temperature
does not occur monotonically.

The time dependences of the elements of the internal stress tensor of the system also undergo a
qualitative change at times close to $t_{0}(T)$. 
All the curves obtained are well approximated by the same expression:

\begin{equation}
{\qquad \Phi (t)=f_{0}+a \cdot t+b\cdot exp((t-t_{0})/\tau) }
\end{equation}

The parameters $f_{0}, a, b, t_{0}, \tau$ for each characteristic ($\sigma, L$) at different temperatures 
were found by fitting with the use of the nonlinear 
least squares algorithm(NLLS)\cite{17}.

\begin{figure}[tbh]
\begin{center}
\resizebox{0.9\columnwidth}{!}{\includegraphics*[angle=-0]{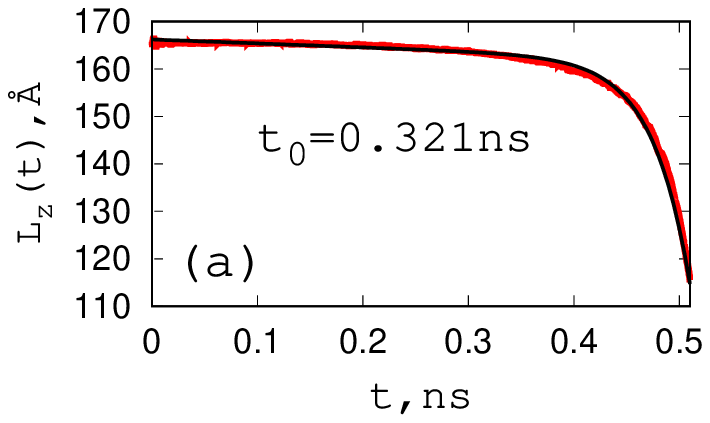}}
\resizebox{0.95\columnwidth}{!}{\includegraphics*[angle=-0]{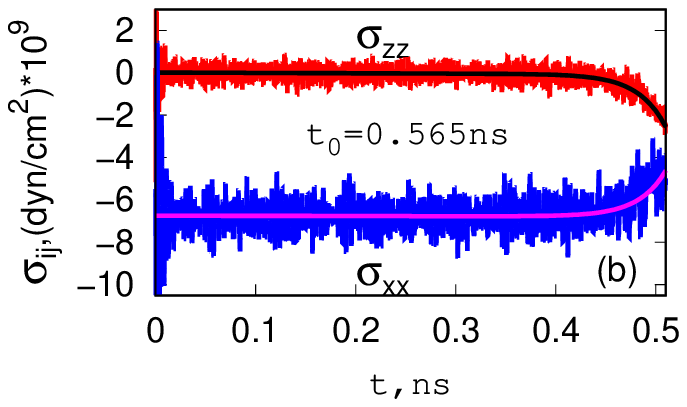}}
\caption{ The time dependences: (a) of the size of the base crystallite along the $z$ axis $L_{z}(t)$ ; 
(b) of the diagonal elements of the stress tensor,
$\sigma_{xx}, \sigma_{zz}$ , at $T = 785K$.}
\label{Fig10}
\end{center}
\end{figure}

\begin{figure}[tbh]
\begin{center}
\resizebox{0.99\columnwidth}{!}{\includegraphics*[angle=-0]{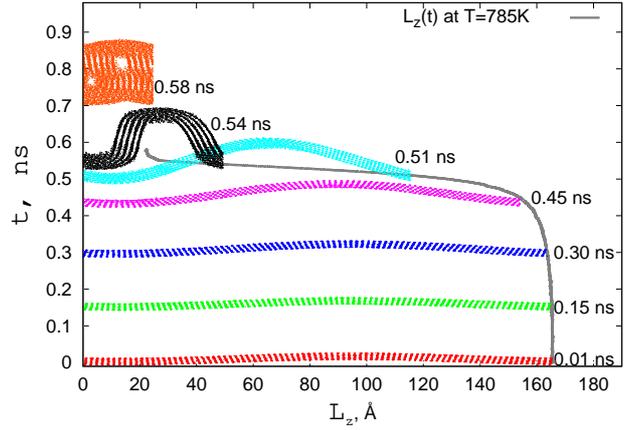}}
\caption{Instantaneous projections of all the film atoms onto the $xz$ plane at various times for $T=785K$.
$L_{z}(t)$ is the size of the base crystallite along the $z$ axis at the same temperature.}
\label{Fig11}
\end{center}
\end{figure}

Since the time characteristics calculated for the considered temperatures $T > 770 K$ are similar, 
we dwell more closely on the results obtained for
a temperature of $785K$.
Figure \ref{Fig10} demonstrates the time dependences of the film size $L_{z}$ (a), and of the diagonal 
elements of the stress tensor $\sigma_{xx}$ and $\sigma_{zz}$
(b) at $T = 785K$. Here, the bold line shows their approximations by formula (4). Only the parameter 
$t_{0}$ determining the moment of qualitative change
of the corresponding characteristics is indicated in the figures:   $t_{0} = 0.321ns$ for $L_{z}$, 
and     $t_{0} = 0.565ns$ for $\sigma$. A comparison of 
these plots shows that, at a given temperature, the system instability is evidenced first on the curve
$L_{z}$, and only later on $\sigma$. 
Figure \ref{Fig11}  shows the instantaneous projections of all the atoms of the calculated cell onto
the $xz$  plane for $T = 785K$ at different  time moments. 
The values of these time moments are shown to the right of the corresponding curves.
The figure demonstrates how the film shape varies with time, as the  instability is approached. 
Note that the instantaneous shape 
of the film coincides with the geometry of the longest low-frequency bending wave, 
whose amplitude increases with time, 
which ensures a linear decrease of $L_{z}$ at times $t < t_{0}$. At  $t > t_{0}$, the process accelerates,
there occurs a rapid increase in the bending wave amplitude, 
the film bends (the wavelength sharply decreases), and the simulation stops.

\begin{figure}[tbh]
\begin{center}
\resizebox{0.99\columnwidth}{!}{\includegraphics*[angle=-0]{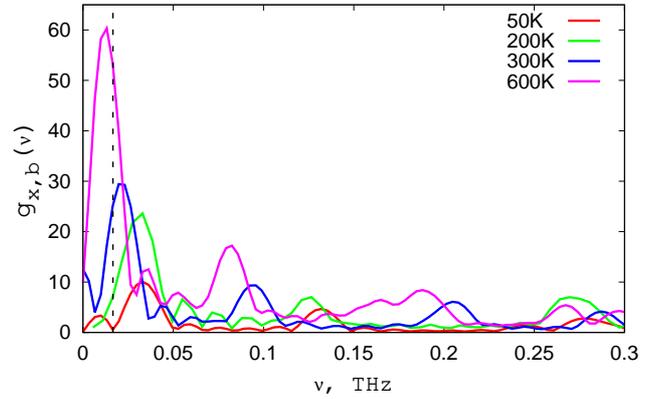}}
\caption{Low-frequency spectra of bending vibrations $g_{x,b}$  of surface atoms at different temperatures.
The dashed vertical line shows the minimum frequency $\nu_{min} = 0.0166 THz$. }
\label{Fig12}
\end{center}
\end{figure}

Let us analyze the changes in the most low-frequency portion of the system spectrum as the temperature 
increases up to $600 K$. 
Figure \ref{Fig12} presents the results of calculation of the frequency spectrum of bending vibrations 
for the temperatures: $50, 200, 300,$ and $600 K$.
The observed spectrum softening with temperature is quite understandable: it is associated with a decrease
in the velocity 
of the transverse sound wave in a bulk sample (see Formula (1)). Already at $600 K$ the vibration period
of this mode equals 
the computation time ($Dt = 0.6 ns$) and, obviously, continues to increase with temperature. Frequencies less
than $1 / Dt$ cannot be determined 
from the Fourier analysis. The vibrational mode with a period larger than the computation time may be 
called ``retarded mode''. 
Indeed, from the beginning of the calculation the amplitude of this wave continuously grows,
the wavelength $\lambda$ thereby decreases, which results in an increase of the film curvature (see Fig.\ref{Fig11}).
If at a moment $t_{0} < Dt$ the wavelength reaches a certain critical value 
($(L_{z}(0)-L_{z}(t_{0}))/L_{z}(0) \sim 0.01$),
the fluctuation  attractive interaction between atoms located on the opposite slopes of the wave crest becomes possible, 
as evidenced by the appearance of negative values of $\sigma_{zz}$  (see Fig.\ref{Fig10}b). 
Once this process has started, it develops exponentially, 
and the film ``collapses''. The program stops calculating when $L_{z}$ becomes less than three radii of the potential cutoff.

The appearance of a ``retarded  mode'' associated with a long bending wave is a necessary condition 
for the development of film destruction process
in the model under study. Indeed, in the other direction (along the $y$ axis) with periodic boundary 
conditions, the time dependence $L_{y} (t)$ has a peculiarity only at those times when the destruction
process in the $z$ direction has already developed, as shown by the dashed line in Fig.\ref{Fig9}.
The geometry of the problem under consideration is such that the maximum wavelength in the $y$ direction 
is $2.8$ times less than along 
the $z$ axis. The period of the corresponding vibrations, proportional to $\lambda^{2}$ (see Formula (1)),
is eight times less than that of a similar 
wave propagating along the $z$ direction, so it cannot become ``retarded'' at any of the temperatures considered.

Thus, three stages can be distinguished in the MD simulation of the thin-film destruction process.

1. With increasing temperature,  the there forms a ``retarded'' bending mode, the vibration period of which is greater than the time
of the steady state existence ($0.6 ns$).

2. A constant in time increase of the amplitude of this mode and, correspondingly, of the film curvature to a critical value, 
when fluctuation attractive interaction between atoms located on the opposite slopes of the wave crest becomes possible.

3. This process develops exponentially, leading to the film destruction.

\section{Conclusion}

The molecular dynamics method is used to study the lattice stability and vibrational properties of a thin 
(5 atomic layers) $fcc$ aluminum film.  
The problem of testing the MMD calculation results in the absence of experimental data for thin films is 
solved by comparing the calculated
dispersion law with the analytical results obtained in the elasticity theory, which requires a special 
attention to be paid to the low-frequency 
portion of the vibrational spectrum of the system. An original approach is proposed, which allows one to 
calculate the dispersion law of long-wave 
phonons in thin films using the molecular dynamics method. The approach is based on a preliminary symmetry
analysis of the vibrations of the film atoms, which enables one to calculate independently the vibrational
spectrum for the symmetrized coordinates corresponding 
to the bending and compression vibrations of a the film. This makes it possible to unambiguously establish 
to which branch of the phonon spectrum 
the considered vibrations with a given frequency $\nu$ belong, and to find the wavelength $\lambda$ of the 
corresponding phonon.
The obtained values of the low-frequency dispersion spectrum are in good agreement with the analytical 
results from the theory of elasticity.
The temperature range ($0 - 600 K$) , in which the film remains stable
for a time of $0.6 ns$ is determined, and 
a frequency analysis of the vibrational spectrum, starting from the minimum frequency $\nu_{min} = 0.0166 THz$ is performed. 
The obtained temperature dependences of the low-frequency spectrum, the changes in the film size along the $z$ axis, 
and in the film shape make it  possible to propose a scenario for the development of the system instability.

It is shown that at the initial stage of film destruction, the main role is played by low-frequency bending vibrations. 
A linear decrease in the size of the base crystallite is due to the increasing bending wave amplitude, 
which is maximum for the chosen problem geometry. 
With increasing amplitude of this wave, the film curvature grows to a critical value, at which the element $\sigma_{zz}$ 
of the internal stress tensor becomes negative, which points to the appearance of fluctuation attractive 
interaction along the $z$ axis between atoms from the opposite slopes of the wave crest, 
following which the process develops exponentially, leading to the film destruction.

\section{Acknowledgments}
 
The work was carried out within the framework of the research topic of the Udmurt Federal Research Center 
of Ural Branch of the Russian Academy of Sciences 
``Theoretical studies of electronic, magnetic, lattice and transport properties of layered and nanostructured systems''
 AAAA17-117022250041-7.

\end{document}